\documentclass[11pt]{article}

   \usepackage{flushend}
 \usepackage{hyperref}
\usepackage{url}
 \usepackage{xcolor,color,url,eurosym,graphicx,amssymb,amsfonts,amsmath,algorithm,algorithmic,bbm,xspace}

  \usepackage{dsfont}
\usepackage{geometry}
 \usepackage{thm-restate}

\newcommand{\vol}[1]{\text{vol}(#1)}

\newcommand{\Abs}[1]{{\left|{#1}\right|}}
\newcommand{\beql}[1]{\begin{equation}\label{#1}}
\newcommand{\eeq}{\end{equation}}

\newcommand{\nlap}{$\mathcal{L}$\xspace}

\newcommand{\field}[1]{\mathbb{#1}} 
\newcommand{\One}[1]{\ensuremath{{\mathbf 1}\left(#1\right)}}
\newcommand{\Prob}[1]{\ensuremath{{\bf{Pr}}\left[{#1}\right]}}
\newcommand{\Mean}[1]{\ensuremath{{\mathbb E}\left[{#1}\right]}}

\newcommand{\NPhard}{{\ensuremath{\mathbf{NP}}-hard}\xspace}

\newcommand{\whp}{\textit{whp}\xspace}

\newcommand{\spara}[1]{\smallskip\noindent{\bf #1}}

\newtheorem{theorem}{Theorem}

\newtheorem{lemma}{Lemma}

\newtheorem{definition}{Definition}

\newcommand{\hide}[1]{}
\newcommand{\squishlist}{
 \begin{list}{$\bullet$}
  {  \setlength{\itemsep}{0pt}
     \setlength{\parsep}{3pt}
     \setlength{\topsep}{3pt}
     \setlength{\partopsep}{0pt}
     \setlength{\leftmargin}{2em}
     \setlength{\labelwidth}{1.5em}
     \setlength{\labelsep}{0.5em}
} }
\newcommand{\squishlisttight}{
 \begin{list}{$\bullet$}
  { \setlength{\itemsep}{0pt}
    \setlength{\parsep}{0pt}
    \setlength{\topsep}{0pt}
    \setlength{\partopsep}{0pt}
    \setlength{\leftmargin}{2em}
    \setlength{\labelwidth}{1.5em}
    \setlength{\labelsep}{0.5em}
} }

\newcommand{\squishdesc}{
 \begin{list}{}
  {  \setlength{\itemsep}{0pt}
     \setlength{\parsep}{3pt}
     \setlength{\topsep}{3pt}
     \setlength{\partopsep}{0pt}
     \setlength{\leftmargin}{1em}
     \setlength{\labelwidth}{1.5em}
     \setlength{\labelsep}{0.5em}
} }

\newcommand{\squishend}{
  \end{list}
}

\newcommand{\squishlistt}{
 \begin{list}{---}
  {  \setlength{\itemsep}{0pt}
     \setlength{\parsep}{3pt}
     \setlength{\topsep}{3pt}
     \setlength{\partopsep}{0pt}
     \setlength{\leftmargin}{2em}
     \setlength{\labelwidth}{1.5em}
     \setlength{\labelsep}{0.5em}
} }

\begin{document}

\title{Scalable motif-aware graph clustering}

\author{
Charalampos E. Tsourakakis \\
Boston University, Harvard University \\ 
babis@seas.harvard.edu \\
\and
Jakub Pachocki \\  
Carnegie Mellon University \\ 
pachocki@cs.cmu.edu 
\and
Michael Mitzenmacher \\ 
Harvard University \\ 
michaelm@seas.harvard.edu 
}

\maketitle

\begin{abstract}
We develop new methods based on graph motifs for graph clustering,
allowing more efficient detection of communities within networks.  We
focus on triangles within graphs, but our techniques extend to other clique
motifs as well.  Our intuition, which has been suggested but not
formalized similarly in previous works, is that triangles are a better
signature of community than edges.  We therefore generalize the notion
of conductance for a graph to {\em triangle conductance}, where the
edges are weighted according to the number of triangles containing the
edge.  This methodology allows us to develop variations of several
existing clustering techniques, including spectral clustering, that minimize triangles split by the cluster instead of edges cut by the cluster.
We provide theoretical results in a planted partition model to demonstrate the potential for triangle conductance in clustering problems.  We then show experimentally the effectiveness of our methods to multiple applications in machine learning and graph mining.  
\end{abstract}

\maketitle

\section{Introduction}
\label{sec:introduction}
Our work is motivated by the following question: {\em how can we
effectively leverage higher-level graph structures, or motifs, for
better clustering and community detection in graph structures}?
Network motifs are basic interaction patterns that recur throughout
networks, much more often than in random networks.  We focus here on triangle subgraphs, which have often been suggested as being stronger signals of community structure than edges alone
\cite{watts1998collective}.  The use of motifs has been leveraged already in the context of dense subgraph discovery \cite{Gionis:2015:DSD:2783258.2789987}, see \cite{mitzenmacher2015scalable,tsourakakis2015kclique}.
For example, social networks tend to be
abundant in triangles, since typically friends of friends tend to
become friends themselves \cite{wasserman1994social}. Triangles are
also important motifs in brain networks \cite{sporns2004motifs}. In
other networks, such as
gene regulation networks, feed-forward loops and bi-fans are known to
be significant patterns of interconnection \cite{milo2002network}, 
but our techniques extend to other such motifs as well.  Despite the
intuition that triangles or other structures may be important for
clustering and related graph problems \cite{benson2015tensor,klymko2014using,satuluri2011local}, there appears to be a gap in
terms of useful formalizations of this idea.  Our main contribution is
a natural and simple formal framework based on generalizing conductance
and related notions such as graph expansion, based on reweighting edges according to the number of triangles that contain the edge.   

\begin{figure*}[htp]
\centering
\begin{tabular}{@{}c@{}@{\ }c@{}@{\ }c@{}} \includegraphics[width=0.33\textwidth]{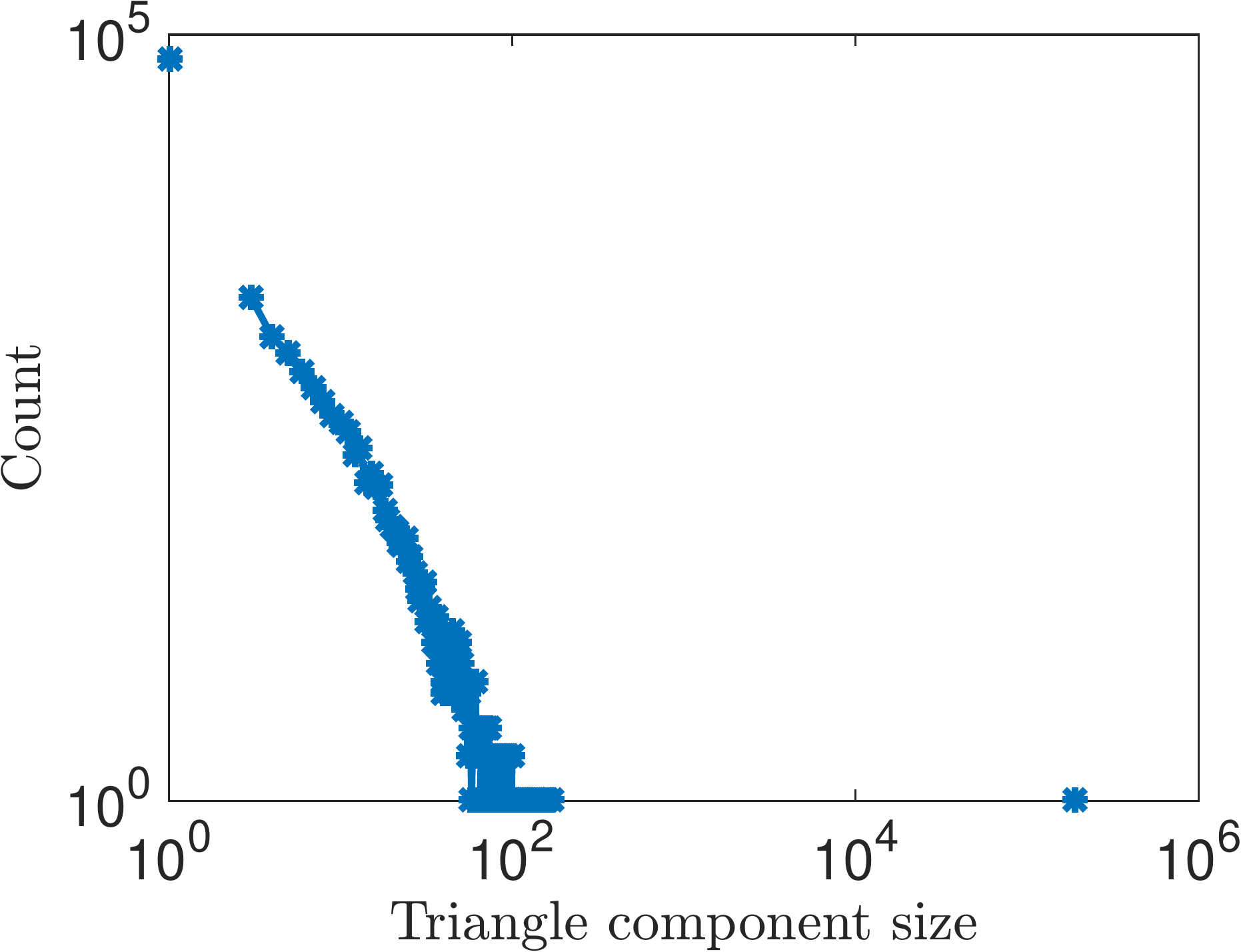} \hspace{10mm}&
\includegraphics[width=0.33\textwidth]{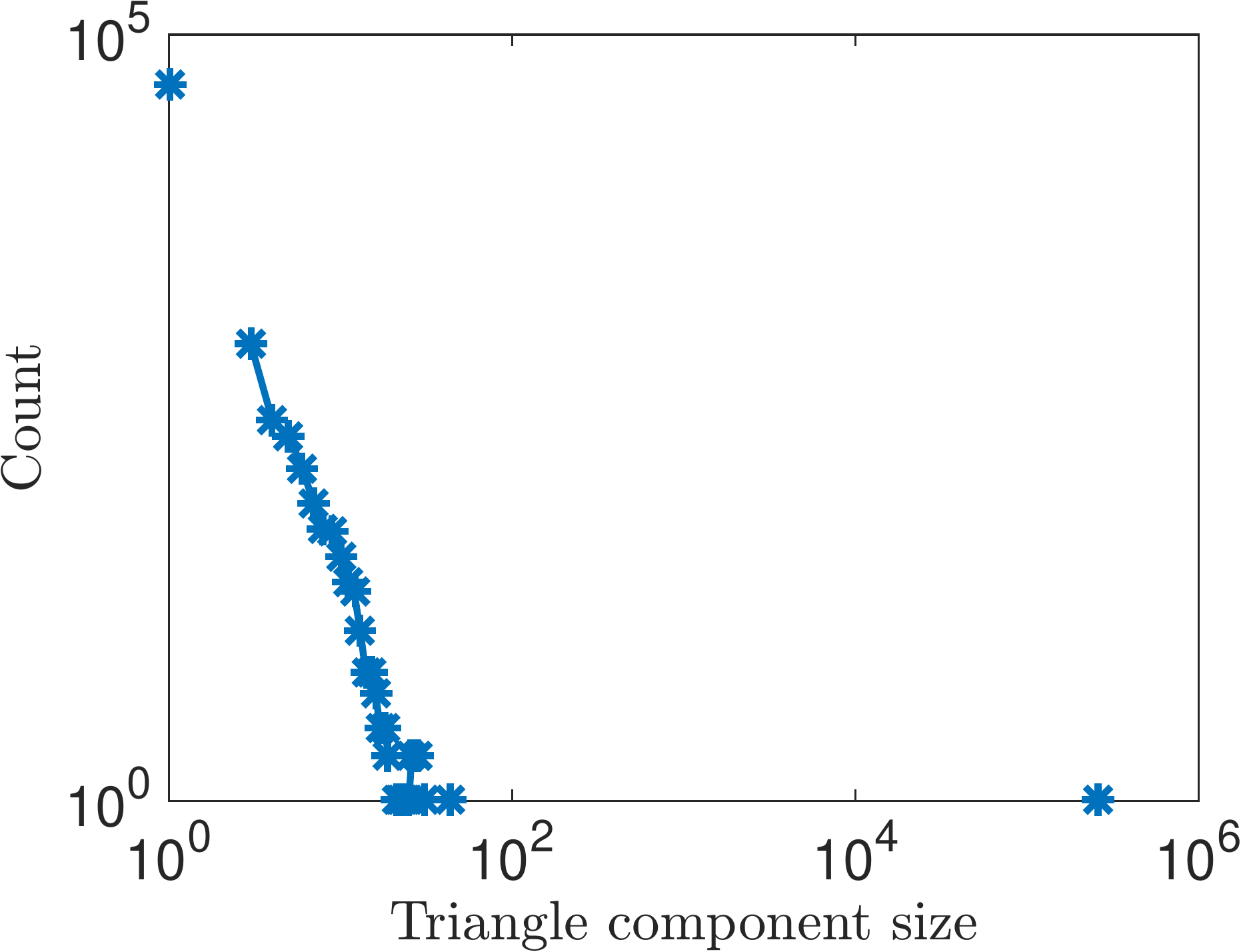} \hspace{2mm} & \includegraphics[width=0.33\textwidth]{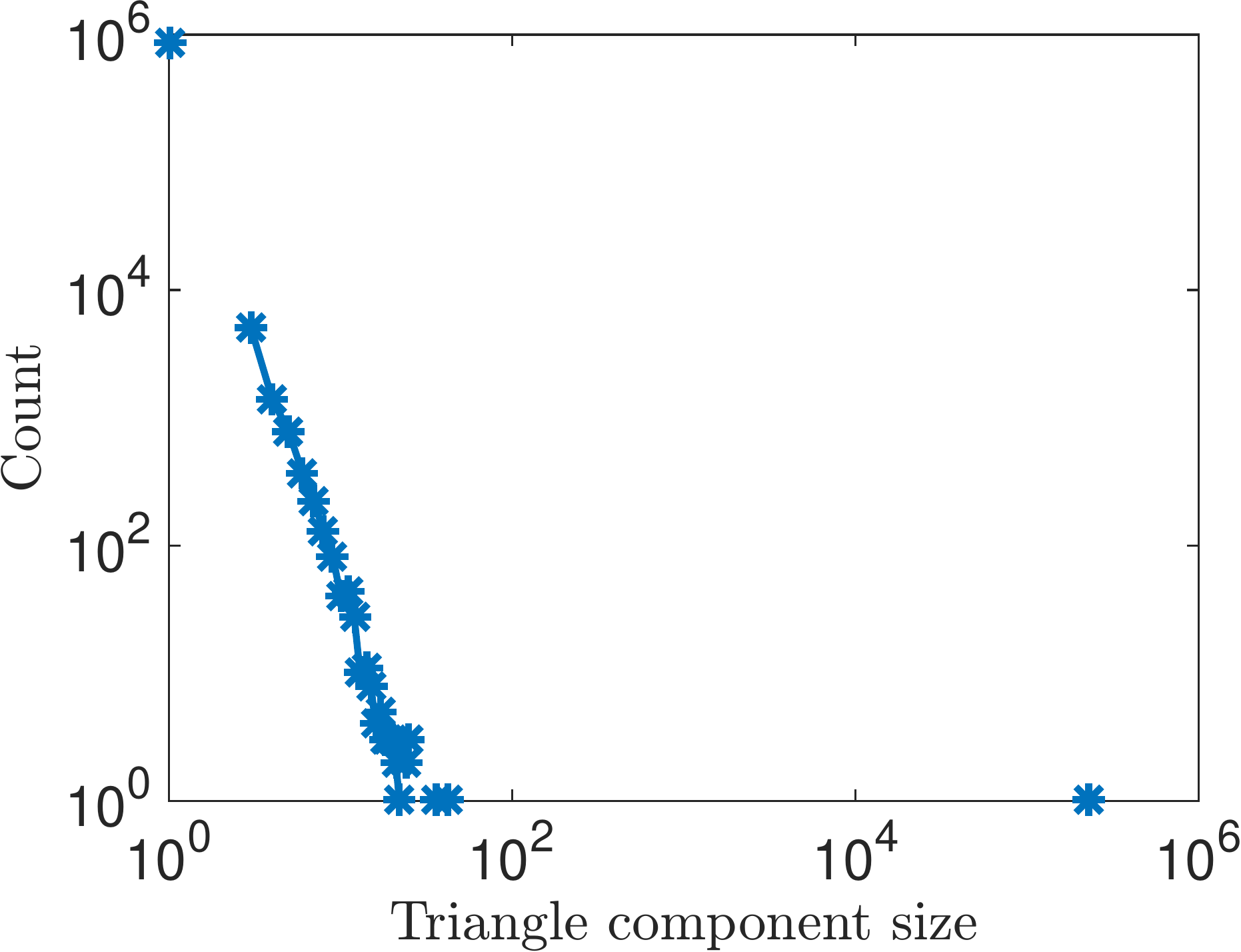} \\
\end{tabular}
\caption{\label{fig:components1} Number of connected components versus  size after reweighing each edge with triangle counts for (a) Amazon, (b) DBLP, and (c) Youtube. The original graphs consist of a single connected component.} 
\vspace{-4mm}
\end{figure*}

\spara{Remark.} Recently, Benson, Gleich, and Leskovec published an article in {\em Science} \cite{benson2016higher} that proposes the same reweighting framework as ours. Our work \cite{tsourakakis2016scalable} and the  {\em Science} paper \cite{benson2016higher}  appeared independently at the same time and share the algorithmic  contribution of performing efficiently motif-based clustering on the input graph without constructing a hypergraph whose hyperedges correspond to  motifs.  In this paper, we have decided to focus on  important contributions of our work that do not appear in \cite{benson2016higher}: a random walk interpretation of the graph reweighting scheme, that provides a principled approach to define the notion of conductance for other motifs;  the framework of motif-based graph expanders that provides the theoretical foundations for motif-based graph clustering; our results on the planted partition model; the introduction of a natural  heuristic that outperforms a wide variety of popular graph community detection methods, both in terms of output quality and run times;  and an experimental evaluation on real-world networks with ground-truth communities.

\spara{Contributions.} Specifically, our contributions are summarized as follows: 

\squishlist
 
\smallskip

\item We formalize intuitions and heuristics in prior work by studying {\em triangle conductance}, 
a variation of graph conductance based on triangles.  Our definitions generalize to other
motifs, but here we focus on triangles. In contrast to prior work  \cite{benson2015tensor,benson2016higher}, we relate the notion of
triangle conductance to appropriate random walks on the graph and to a generalization of graph expansion based on triangles instead of edges. When at node $u$  we choose a triangle that $u$ participates in uniformly at random and then choose an endpoint of that triangle, other than $u$, uniformly at random. We differentiate our new concepts by for example showing that an
expander graph \cite{alon1987better} is not necessarily a triangle
expander and vice versa. 

\smallskip
\item We provide approximation algorithms for a generalization of the well-studied sparsest cut problem \cite{vazirani2013approximation},
where the goal now is to minimize the number of triangles cut by a partition.  We present this part of our work briefly as it coincides with the algorithmic contribution of the {\em Science} paper \cite{benson2016higher}.

\smallskip

\item We study our reweighting algorithm in the planted partition model, where we provide tight theoretical guarantees on its ability to recover the true graph partition with high probability\footnote{An event $A_n$ holds with high probability (whp) if $\lim_{n \rightarrow +\infty} \Prob{A_n}=1$.}.

\smallskip

\item  We propose a highly effective heuristic method for detecting communities.  Specifically,  using publicly available datasets where ground-truth is available, we verify the effectiveness of our framework, and show it takes orders of magnitude less time and obtains similar performance to the best performing competitor Markov clustering (MCL) \cite{dongen2000graph}. 

\squishend

Before beginning, we show that our scheme reweighting edges by triangle counts provides significant insights on the community structure of real-world networks.  Surprisingly, in many real-world networks we find this simple step immediately disconnects the graph into numerous non-trivial connected components, that we refer as triangle components. Figure~\ref{fig:components1} shows the distribution of triangle components for the {\sc Amazon}, {\sc DBLP}, and {\sc Youtube} networks (see Table~\ref{tab:datasets} for a detailed description).  Our findings are consistent across all of them: there exists one giant triangle component and then a large number of triangle components with up to few hundreds of nodes.   (Trivially all degree one nodes in the original graph become isolated components.)   

These findings agree with the ``jellyfish'' or ``octopus'' model \cite{tauro2001simple}, according to which most networks have a giant ``core'' with a large number of relatively small ``whiskers'' dangling around. Furthermore, our findings agree with the findings of  \cite{leskovec2008statistical} that claim that  communities have size up to roughly 100 nodes.  Our findings  show additionally to \cite{leskovec2008statistical} that no triangles are split between whiskers and the rest of the graph.  
We generalize this idea for our clustering results and experiments.  

\spara{Roadmap.}  Section~\ref{sec:related}  briefly presents  related work. Section~\ref{sec:proposed} presents our algorithmic contributions, and Section~\ref{sec:exp} studies the performance of our proposed methods and various competitors on graphs with community ground-truth available. Section~\ref{sec:foundations} sets the theoretical foundations for motif-based community detection.

\spara{Notation.} We use the following notation throughout the paper. Let $G(V,E,w)$ be an undirected graph with non-negative weights;  we also use $G(V,E)$ for unweighted graphs.  The weighted degree $deg(u)$ of a node $u \in V$ is equal to $deg(i) = \sum_{j \in V} w(i,j)$. 
For a set of nodes $S \subseteq V$ we define $w(S:\bar{S}) = \sum\limits_{i \in S, j \in \bar{S}} w(i,j)$  as the total weight of the edges leaving $S$.  Also let $\vol{S}=\sum\limits_{i \in S} deg(i)$ be the volume of $S$.  For the case of unweighted graphs, we denote $w(S:\bar{S})$ as $e(S:\bar{S})$ for clarity, and we define  $t(u), t(u,v)$ as the total number of triangles that contain node $u$ and edge $(u,v) \in E(G)$ respectively. Notice for unweighted graphs  graphs,  $\vol{S}=2e(S)+e(S:\bar{S})$ where $e(S)$ is the number of edges induced by $S$.

\section{Related Work}
\label{sec:related}
\spara{Communities.} Intuitively, a community is a set of nodes with more and/or better intra- than inter-connections. There are different approaches to defining the notion of a community that lead to different mathematical formalizations. For instance, the notion of modularity captures the difference between the connectivity structure of a set of nodes compared the expected structure if edges in the graph were distributed at random \cite{newman2006modularity}.  
Conductance is one of the most popular measures used in community detection  \cite{fortunato2010community,leskovec2008statistical,schaeffer2007graph,yang2015defining}.  It quantifies the intuition that the total weight of edges leaving the community should be relatively small compared to the internal weight. 
 It is worth outlining that this intuition is not always true \cite{abrahao2012separability}. Specifically, there exist networks with communities whose outgoing number of edges  
{\em is not} small compared to the number of internal edges. 
The notions of $k$-clique communities \cite{derenyi2005clique}, i.e., the union of all cliques of size $k$ that can be reached through adjacent  $k$-cliques that share $k-1$ nodes,  and $(\alpha,\beta)$-communities \cite{mishra2008finding} have been proposed to tackle communities whose outgoing number of edges  
{\em is not} small compared to the number of internal edges. 

We formally define graph conductance. For any set $S \subseteq V$ we define its expansion, also known as conductance, by 

$$\phi(S) = \frac{w(S:\bar{S}) }{ \min(\vol{S}, \vol{\bar{S}} ) }.$$ 

\noindent The edge expansion of the graph, also known as graph conductance, is defined as $\phi(G) = \min_S \phi(S)$.

Given a connected graph $G$, finding cuts with minimum conductance 
is \NPhard. A lot of work has focused on developing approximation algorithms \cite{arora2009expander,leighton1999multicommodity,alon1985lambda1}. As noted in numerous works, cf.\cite{leskovec2008statistical}, spectral clustering is considered to be the most practical approach.  

\spara{Spectral clustering.} Cheeger's inequality  establishes a bound on edge expansion via the spectrum of the normalized Laplacian matrix representation of the graph. Specifically, let  $A$ be the adjacency matrix of $G$, and $D$ a diagonal matrix containing the weighted degrees in its diagonal. The combinatorial Laplacian 
is defined as  $L = D-A$.  The normalized Laplacian 
is  \nlap$=D^{-1/2}L D^{-1/2}$.  It is well-known that the multiplicity  of the zero eigenvalue of \nlap equals the number 
of connected components of $G$. Let us assume without any loss of generality that $G$ is connected, hence only one eigenvalue of \nlap equals 0. 
The following theorem forms the basis of spectral graph theory. 
 \vspace{-1mm}
\begin{theorem}[Discrete Cheeger's inequality \cite{alon1985lambda1}]
Given a weighted undirected graph $G(V,E,w)$ and its normalized Laplacian matrix \nlap, let the eigenvalues of \nlap be 
$0 = \lambda_1 \leq \lambda_2 \leq \ldots \leq \lambda_n \leq 2 $.
Then $\frac{\lambda_2}{2} \leq \phi(G) \leq \sqrt{2\lambda_2}$.

\hide{ 
\beql{cheeger}
\frac{\lambda_2}{2} \leq \phi(G) \leq \sqrt{2\lambda_2}.
\eeq
}
\end{theorem}
\vspace{-1mm}

\noindent Cheeger's inequality is the basis of spectral clustering \cite{ng2002spectral,von2007tutorial}.   While there exist various versions of spectral clustering, its basic form consists of the following three steps: 
(i)  Compute the eigenvector $x$ of $\lambda_2$, 
and sort its entries so that $x_1\leq x_2 \leq \ldots \leq x_n$. 
(ii) Consider subsets $S_i = \{ x_1,\ldots, x_i\}$. (iii) Output $S = \text{arg}\min \phi(S_i)$. The output $S$ has conductance $\phi(S) \leq \sqrt{ 2\lambda_2}$.  
Cheeger's inequality has recently been  generalized to hypergraphs by Louis \cite{louis2014hypergraph}.

\spara{Expander graphs.}  Intuitively an expander is a graph that contains no set $S$ with low conductance.  Expander graphs  with constant degree play an important role in a wide variety of applications, including coding theory and hashing. The interested reader may read the excellent monograph of Hoory, Linial, and Widgerson for more details \cite{hoory2006expander}.  The formal definition follows.
 
\begin{definition}[Expander]
A graph $G(V,E,w)$ where $w:E\rightarrow \field{R}^+$ is an expander if all subsets $S \subseteq V$ with $|S|\leq 0.5 n$ 
have  edge expansion $\phi(S)=\Theta(1)$.
\end{definition}

 \spara{Triangle biased random walks.} Motifs, and specifically triangles, have been used in random walks, e.g., \cite{benson2015tensor,backstrom2011network}. For example, Backstrom and Kleinberg \cite{backstrom2011network} used weighted triangle closing walks as follows: when a random walk is at node $u$ and considers which neighbor of $u$ it should choose, it remembers the previous node in the walk $s$. If $(s,v)$ is an edge, then the walk is biased towards $v$. According to their findings, this is a successful heuristic for detecting better quality clusters compared to standard random walks.

\section{Algorithms}
\label{sec:proposed}

\subsection{Theoretical Framework}
\label{conductance}

\spara{Triangle Conductance.} Let $G(V,E)$ be an unweighted, undirected graph, and set $\text{vol}_3(S) = \sum_{v \in S} t(v)$. From now on, we denote $\vol{S}$ as $\text{vol}_{2}(S)$ in order to distinguish  $\text{vol}_{2}$ and $\text{vol}_{3}$.  Also, for a set $S \subseteq V$, define $t_i(S)$ to be the number of triangles with exactly $i$ vertices in $S$. By double counting we obtain $ \text{vol}_3(S) = 3t_3(S) + 2t_2(S) + t_1(S)$. Consider the following  biased random walk that utilizes the intuition that triangles play an important role in community detection. When at node $u$ the random walk chooses a neighbor  $v \in N(u)$ with probability proportional to   $t(u,v)$. Equivalently, when at node $u$  we choose a triangle that $u$ participates in uniformly at random and then choose an endpoint of that triangle, other than $u$, uniformly at random.  Notice that if the random walk starts at a vertex $u$ that does not participate in any triangles, i.e., $t(u)=0$, then the random walk stays at $u$. Let $S \subseteq V$ be any   set of vertices,  and denote by $\phi_3(S)$ the probability of leaving $S$ in one step of the walk conditioned on being at a vertex $u$ chosen from $S$ proportionally to the number of triangles $t(u)$ it participates in\footnote{To see why  $\phi_3(S)$  equals to the escape probability  notice that  $\sum_{u \in S} \frac{t(u)}{\text{vol}_3(S)} \frac{ 0 \times t_3(u)+ 0.5\times t_2(u) + 1 \times t_1(u)}{t(u)}=\frac{  t_2(S) + t_1(S) }{ \text{vol}_3(S)  }$. Here $t_i(u)$ is the number of triangles with $i$ vertices in $S$ ($u$ included).} .   Then,

$$ \phi_3(S) = \frac{ 2t_2(S) + 2t_1(S) }{ 6t_3(S)+2t_2(S)+  2t_2(S) + 2t_1(S) }=\frac{  t_2(S) + t_1(S) }{ \text{vol}_3(S)  }.$$ 

\noindent Clearly $\phi_3(S) \in [0,1]$.  We define the graph triangle conductance   as 

$$ \phi_3(G) = \min\limits_{S \subseteq V} \frac{  t_2(S) + t_1(S) }{ \min{( \text{vol}_3(S),\text{vol}_3(\bar{S}))} }.$$ 

\noindent Notice that the denominator is set to the minimum of the triangle volumes because of the symmetry  $t_2(S) + t_1(S)= t_2(\bar{S}) + t_1(\bar{S})$.

\subsection{Triangle Spectral Clustering}
\label{cheeger}
We provide an efficient approximation algorithm for the triangle conductance problem.Notice that this is essentially a hypergraph problem where each hyperedge corresponds to a triangle. 
For a given input graph $G(V,E)$ with a set of triangles 
$\mathcal{T}_G  \subseteq {[n] \choose 3}$, 
define the 3-uniform   hypergraph $\mathcal{H}(V,E_{\mathcal{H}})$, 
where each hyperedge $e \in E_{\mathcal{H}}$ corresponds to 
a triangle $u,v,w \in \mathcal{T}_G$. Consider any cut $(S:\bar{S})$ in $G$ and $\mathcal{H}$. The number of triangles $t(S:\bar{S})$ that go across the cut $(S:\bar{S})$ in $G$ is equal to the number of hyperedges going across $(S:\bar{S})$
 in $\mathcal{H}$.  However, creating $\mathcal{H}$ and then using state-of-the-art 
 semidefinite programming techniques for spectral clustering in \cite{louis2014hypergraph} is computationally expensive.
  Our main theoretical result overlaps with the algorithmic contribution of \cite{benson2016higher}, and is stated here without proof, for completeness reasons. Our result provides an efficient way to perform triangle  spectral clustering.  The interested reader can read our proof on arxiv \cite{tsourakakis2016scalable}.

\begin{theorem} 
Given an undirected, connected graph $G(V,E)$, let $w:E \rightarrow \field{R}^+$ be the weight function that assigns to each edge $e$ weight $w(e)$ equal to the number of triangles $t(e)$ that $e$ is contained. Let $H(V,E,w)$ be the weighted version of $G$. 
Let the eigenvalues of $\mathcal{L}_H$ be 
$0 = \lambda_1 < \lambda_2 \leq \ldots \leq \lambda_n \leq 2 $.
 Then Cheeger's clustering algorithm on $H(V,E,w)$ outputs a cut $(S:\bar{S})$ such that 

\beql{cheeger-triangles}
\frac{\lambda_2(H)}{2} \leq \phi_3(G) \leq \sqrt{2\lambda_2(H)}.
\eeq
\end{theorem}

\spara{Quadratic form for triangle clustering.}   
We define for each triangle $\Delta(u,v,w)$ a $n\times n$ positive semidefinite matrix $L_{\Delta(u,v,w)}$ that is zero except at the intersection of rows and columns indexed by $u, v, w$.   The non-zero entries are $L_{\Delta(u,v,w)}(i,i)=2$ for $i \in \{u,v,w\}$, 
and $L_{\Delta(u,v,w)}(i,j) = -1$ for $i\neq j, i,j \in \{u,v,w\}$. In other words, the $3\times 3$ non-zero sub-matrix of $L_{\Delta(u,v,w)}$ indexed by $u,v,w$ equals

\[  L_{\Delta(u,v,w)}= \left( \begin{array}{ccc}
2 & -1 & -1 \\
-1 & 2 & -1 \\
-1 & -1 & 2 \end{array} \right)\] 

\noindent   Let $x \in \{0,1\}^n$ be the indicator vector of a cut $(S,V\backslash S)$. Specifically, let $x(u)=1$ if and only if 
$u \in S$.  Consider the positive semidefinite matrix $ Q= \sum_{\Delta(u,v,w)} L_{\Delta(u,v,w)}$.  Notice that

\begin{align*}
x^T Q x &= \sum_{\Delta(u,v,w)} \Big( (x_u -x_v)^2+ (x_u -x_w)^2+(x_w -x_v)^2 \Big)  = 2t_2(S)+2t_1(S).
\end{align*} 

\noindent The spectral approach has been evaluated in \cite{tsourakakis2016scalable}, and extensively in  \cite{benson2016higher}, and  has been shown to be very effective in revealing successfully communities in a wide variety of applications. In the next section, we propose TECTONIC, a significantly faster method compared to spectral clustering that produces high quality output as we will see in Section~\ref{sec:exp}.

\subsection{Proposed Method: TECTONIC}
\label{heuristic}
In Section~\ref{sec:introduction} we saw that  reweighting each  edge $(u,v) \in E(G)$ of the graph with weights equal to the triangle count $t(u,v)$ results in disconnecting the graph into multiple connected components. But do these components correlate at all with communities? As we will see in detail in Section~\ref{sec:exp}, they do correlate but there is room for improvement. The main issue with the simple reweighting scheme is that it does not handle well  imbalance, i.e., the existence of communities with different numbers of nodes. Our proposed method {\sc Tectonic} (Triangle Connected Component Clustering, see Algorithm~\ref{alg:heuristic}) deals with imbalance by normalizing the triangle weight $t(u,v)$ by the sum of degrees $deg(u)+deg(v)$.  Then, it removes all edges with weight less than a predefined threshold $\theta$. It is  worth outlining that {\sc Tectonic} is amenable to distributed implementation as it relies simply on triangle counting and thresholding.  

\begin{algorithm}[t]
\caption{\label{alg:heuristic} {\sc Tectonic} } 
 \begin{algorithmic} 
\REQUIRE Undirected, unweighted, connected graph $G(V,E)$
\REQUIRE Threshold $\theta>0$ 
\STATE Count $t(u,v)$ for each  $(u,v) \in E$  
\STATE  Reweight each edge $(u,v) \in E$ by $w(u,v) \leftarrow \frac{t(u,v)}{deg(u)+deg(v)}$
\STATE Remove all edges $(u,v)$ with weight $w(u,v) < \theta$
\STATE Output the resulting connected components 
\end{algorithmic}
\end{algorithm}

Our heuristic normalization scheme is inspired by the following observation. Let $\theta = \frac{1}{2}\Big( 1- \frac{\theta'}{deg(u)+deg(v)}\Big)$. Then two neighboring nodes $u,v$ in $G$ become disconnected after reweighting if and only if

\begin{align*}
\frac{ t(u,v)}{ deg(u)+deg(v) } &< \theta \Leftrightarrow  
\frac{1}{2} (deg(u)+deg(v)-\theta') > t(u,v) \Leftrightarrow \\  
deg(u)+deg(v) - 2t(u,v) &> \theta' \Leftrightarrow    
|N(u) \cup N(v)| - |N(u) \cap N(v)| > \theta' \Leftrightarrow \\ 
dist^2( A^{(u)}, A^{(v)} ) &> \theta', 
\end{align*}

\noindent where $N(u) = \{ v: (u,v) \in E(G)\}$, and $dist( A^{(u)}, A^{(v)} )$ is the Euclidean distance between the $u$-th and $v$-th row of the adjacency matrix representation of $G$.

\section{Experimental results}
\label{sec:exp}
\subsection{Experimental setup} 

Table~\ref{tab:datasets} shows the three networks we use in our experiments together with  the number of nodes $n$ and the number of edges $m$. We use three social and information graphs for which ground-truth about the community structure is available  \cite{snap}. For all datasets we use the top 5000 ground-truth communities, as provided by SNAP.

As our competitors we use a list of popular graph clustering methods: MCL \cite{dongen2000graph}, Infomap \cite{rosvall2008maps},  the Girvan-Newman (GN) algorithm \cite{girvan2002community}, the Louvain method \cite{blondel2008fast}, the Clauset-Newman-Moore (CNM) \cite{clauset2004finding} , Cfinder \cite{adamcsek2006cfinder}, spectral clustering (SC) \cite{ng2002spectral}, and  triangle spectral clustering (tSC) \cite{benson2016higher,tsourakakis2016scalable}.
For the Girvan-Newman algorithm, we use the implementation available at SNAP, and for spectral clustering (SC,tSC) we use the Python sklearn library.  For all other methods we use the original implementations provided by the authors. Methods that had not completed after several hours were stopped. Our code will become available at \url{https://github.com/tsourolampis/tectonic}. Our results were obtained by setting $\theta=0.06$.  We discuss the choice of $\theta$ in the next Section, as a rule of thumb we suggest this value.

We count triangles exactly using {\sc Mace} \cite{maceuno,uno}. All experiments run on a laptop with 1.7 GHz Intel Core i7 processor and 8GB of main memory.  Triangle counting took 
0.56, 1.25 and 6.6 seconds for Amazon, DBLP, and Youtube graphs respectively.  

\begin{table}[!ht]
\begin{center}
\begin{tabular}{|l|c|c|} \hline
Name  & $n$  & $m$  \\ \hline 
{\sc Amazon} &    334\,863 & 925\,872	 \\
{\sc  DBLP }  & 317\,080 & 1\,049\,866 \\
{\sc YouTube}   & 1\,134\,890	& 2\,987\,624	 \\ \hline
\end{tabular}
\end{center}
\caption{\label{tab:datasets} Datasets used in our experiments. }
\end{table}

\subsection{Community detection}

\begin{table*}[!ht]
\begin{center}
\begin{tabular}{|l|c|c|c|c|c|c|c|c|c|}\hline 
Method       &  \multicolumn{3}{|c|}{Amazon}  &  \multicolumn{3}{|c|}{DBLP}  &  \multicolumn{3}{|c|}{YouTube}\\ \hline
            
			 &  \multicolumn{1}{|c|}{p} &  
			 \multicolumn{1}{|c|}{r} &  
			 \multicolumn{1}{|c|}{T} &
			 \multicolumn{1}{|c|}{p} &  
			 \multicolumn{1}{|c|}{r} &  
			 \multicolumn{1}{|c|}{T} &
			 \multicolumn{1}{|c|}{p} &  
			 \multicolumn{1}{|c|}{r} &  
			 \multicolumn{1}{|c|}{T} \\ \hline
MCL &  95.6  &  90.1   &  736.54   &   55.1   & 81.7   & 1\,166  &   39.9 & 60.6      & 19\,187.1     \\ \hline           
Louvaine & 50.0 & 14.7 & 9.00 &  50.20 & 12.13 & 10.38 & 50.13 & 27.55& 55.8  \\ \hline
CFinder  & - & - & $>5h$ &  - & - & $>5h$ & - & - & $>5h$   \\ \hline  
GN & - & - & $>5h$ &  - & - & $>5h$ & - & - & $>5h$   \\ \hline                                                                                                
CNM  & - & - & $>5h$ &  - & - & $>5h$ & - & - & $>5h$    \\ \hline   
Infomap     & 50.0    &  14.8    &     63.0   &  50.16     &  12.13   &   64.0  &    50.00 & 27.6   &      204  \\ \hline
SC   & - & - & $>5h$ &  - & - & $>5h$ & - & - & $>5h$    \\ \hline  
tSC  & - & - & $>5h$ &  - & - & $>5h$ & - & - & $>5h$   \\ \hline   \hline
Thres. 0     & 85.2  & 96.0     & 4.62     &  4.0     &  100.0 & 1.65  & 22.5  & 70.8         &6.92  \\ \hline
Thres. 1       &    94.1 &   81.1 &   4.61   &   12.0   & 91.4   & 1.65   &   36.1 & 59.7  & 6.92      \\ \hline
Thres. 2     &   97.1  &  67.7  & 4.62     & 23.0     & 81.6   &  1.65  &   45.0 & 53.9        & 6.92   \\ \hline
Thres. 3    &  98.0  & 52.4    & 4.62     &  35.7    & 71.4   &  1.65  &  49.6 & 50.3      & 6.93    \\ \hline
TECTONIC        &   94.9  & 91.3    &     4.62  & 48.3     & 79.1   &   1.65  &    66.7 & 43.3 &      6.92       \\ \hline                                                                                            
\end{tabular}
\end{center}
\caption{\label{tab:groundtruth} Average precision (p), average recall (r) over all ground-truth communities, and total run time (T) in seconds for MCL and our method using various threshold values. The run times for our method include   the run time for triangle counting (0.56, 1.25 and 6.6 secs respectively).}
\end{table*}

Table~\ref{tab:groundtruth} shows our experimental findings. For each method we use we report the average precision and recall over all 5\,000 ground-truth communities. We compute the precision and recall of a given partition as follows: for each ground-truth community $S$, we find the  community $S'$ in the partition that has the largest intersection size   with $S$. Then, we compute how  well $S'$ matches $S$ by computing  precision and recall. The overall precision and recall that we report is averaged over all ground-truth communities. Method Thres. 0 refers to just reweighting edges by triangle counts, as in Figure~\ref{fig:components1}. Methods Thres. 1,2,3 take this idea further, by removing edges whose weight is less or equal than 1,2,3 respectively. 
Surprisingly, this simple reweighting reveals a lot about the community structure.  For example, as soon as we add triangle weights the single connected component of Amazon breaks up into 77\,811 components. When we remove all edges whose weight is 1,  we obtain 139\,456 components. Similarly for threshold values 2, 3 we find  199\,693  and 250\,572 connected components. Precision and recall show that these  components correlate well with the ground-truth communities. The same is true for the other two datasets. Analyzing further the  ground-truth communities shows that they typically have low conductance 
$\phi_2$. Therefore,  on these datasets  low values of $\phi_2$ and $\phi_3$ are positively correlated. Nonetheless, reweighting by triangle counts may immediately reveal  the community structure or  lower the conductance further, i.e., $\phi_3(S) < \phi_2(S)$. Even in the latter case, this facilitates the algorithmic discovery of such communities.

\begin{figure}[!ht]
\centering
\begin{tabular}{@{}c@{}@{\ }c@{}@{\ }c@{}} \includegraphics[width=0.33\textwidth]{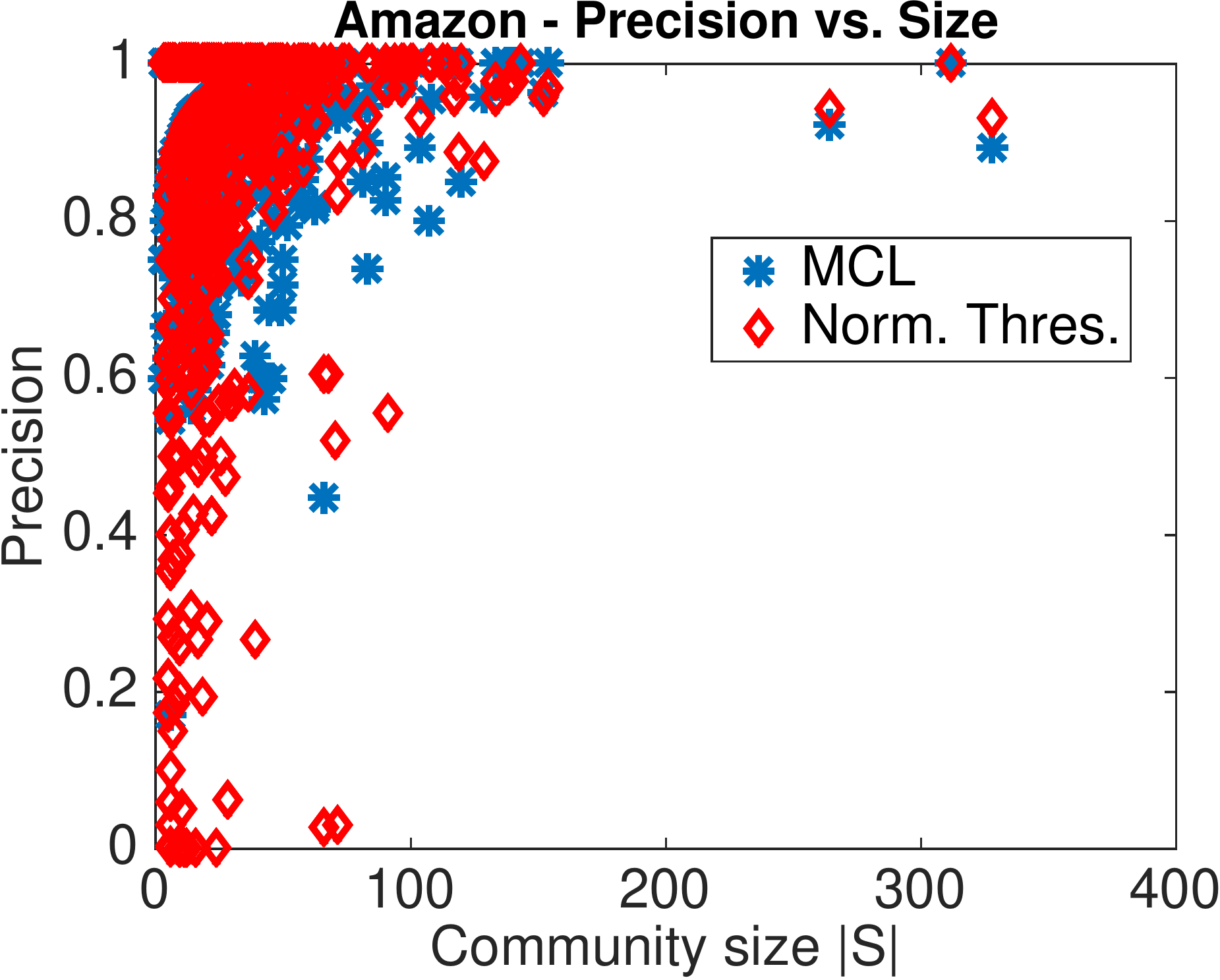} \hspace{3mm} \hspace{2mm} &
\includegraphics[width=0.33\textwidth]{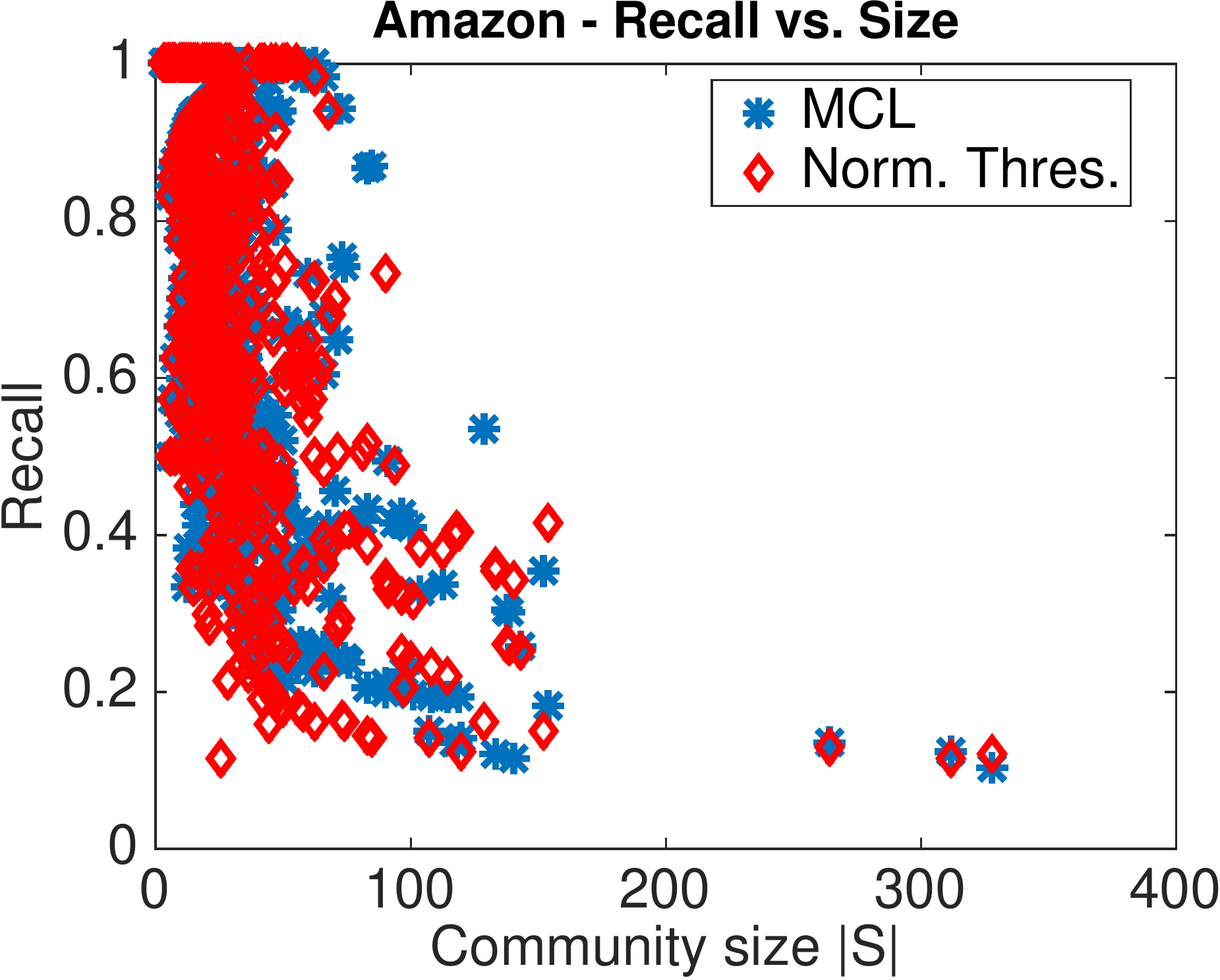} \hspace{5mm} &   \includegraphics[width=0.28\textwidth]{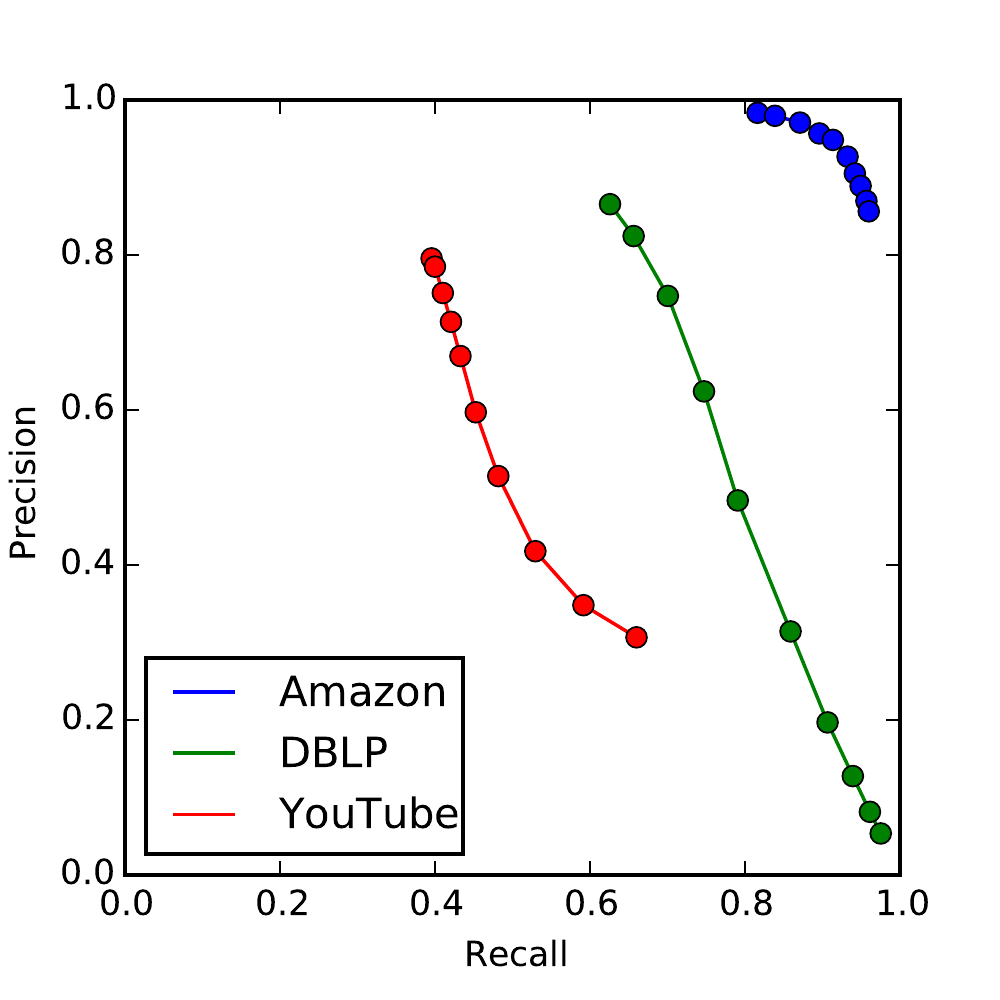} 
\\
(a) & (b) & (c)
\end{tabular}
\caption{\label{fig:amazon} (a) Precision, and (b) Recall vs. ground-truth community size for the Amazon graph using MCL \cite{dongen2000graph}, the best competitor, 
and our method TECTONIC (Norm. Thres.). (c) Precision vs. recall for our method for various threshold values ranging from 0.01 to 0.1 with a step of 0.01. }
\vspace{-4mm}
\end{figure}

In terms of run times, our methods are significantly faster than other methods. At one extreme, CFinder, GN, CNM, SC, tSC do not produce any output after running for at least 5 hours. Actually, GN does not produce any output after running for at least 10 hours.  Louvain is the fastest method among competitors but produces significantly lower quality output compared to MCL. Infomap has a similar behavior to Louvain, but is slightly slower. 
Our method only requires a few seconds, as it only needs to compute the degree sequence, the triangle counts, and the connected components.

\begin{figure*}[!ht]
\centering
\begin{tabular}{@{}c@{}@{\ }c@{}} \includegraphics[width=0.49\textwidth]{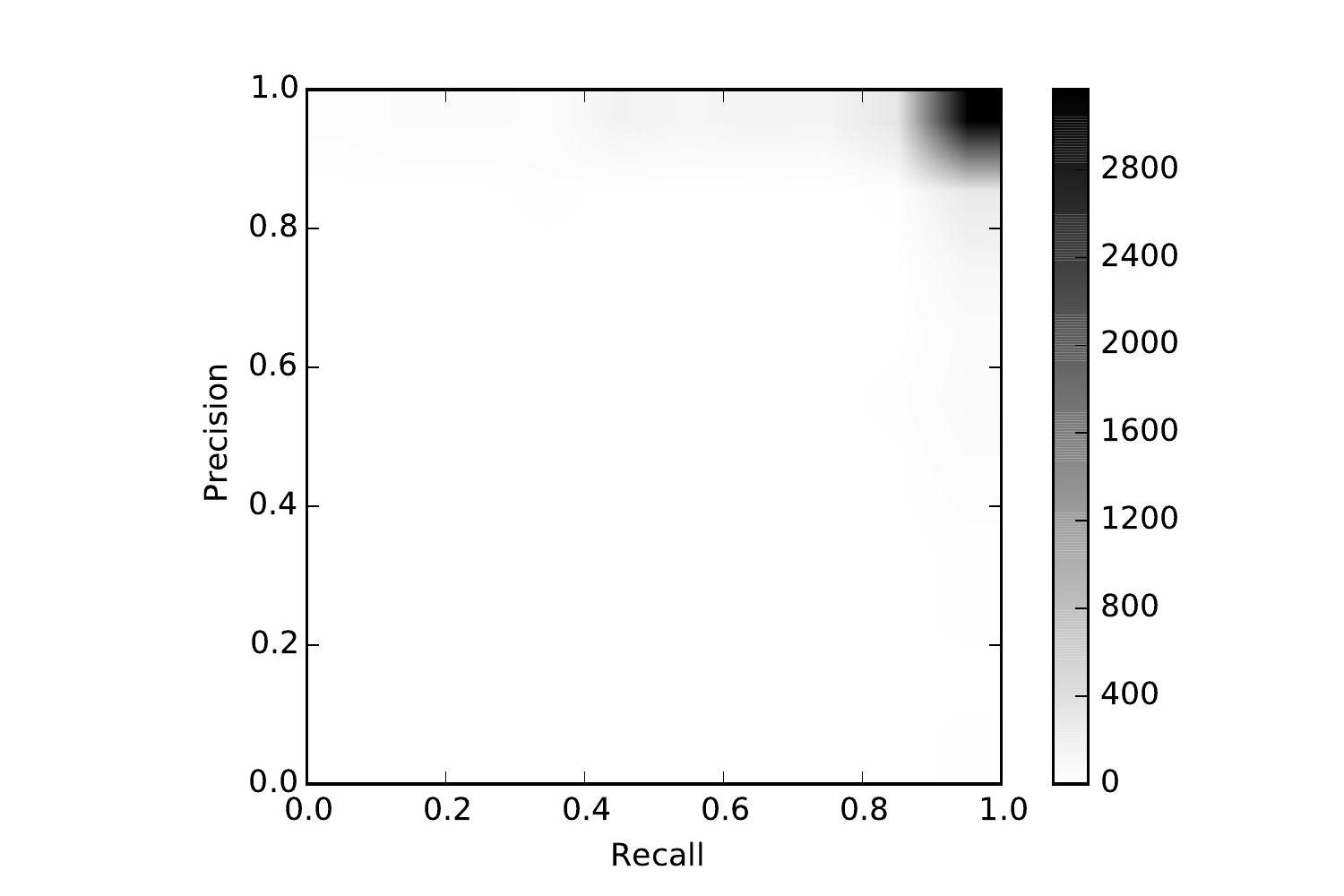} \hspace{3mm} &
\includegraphics[width=0.49\textwidth]{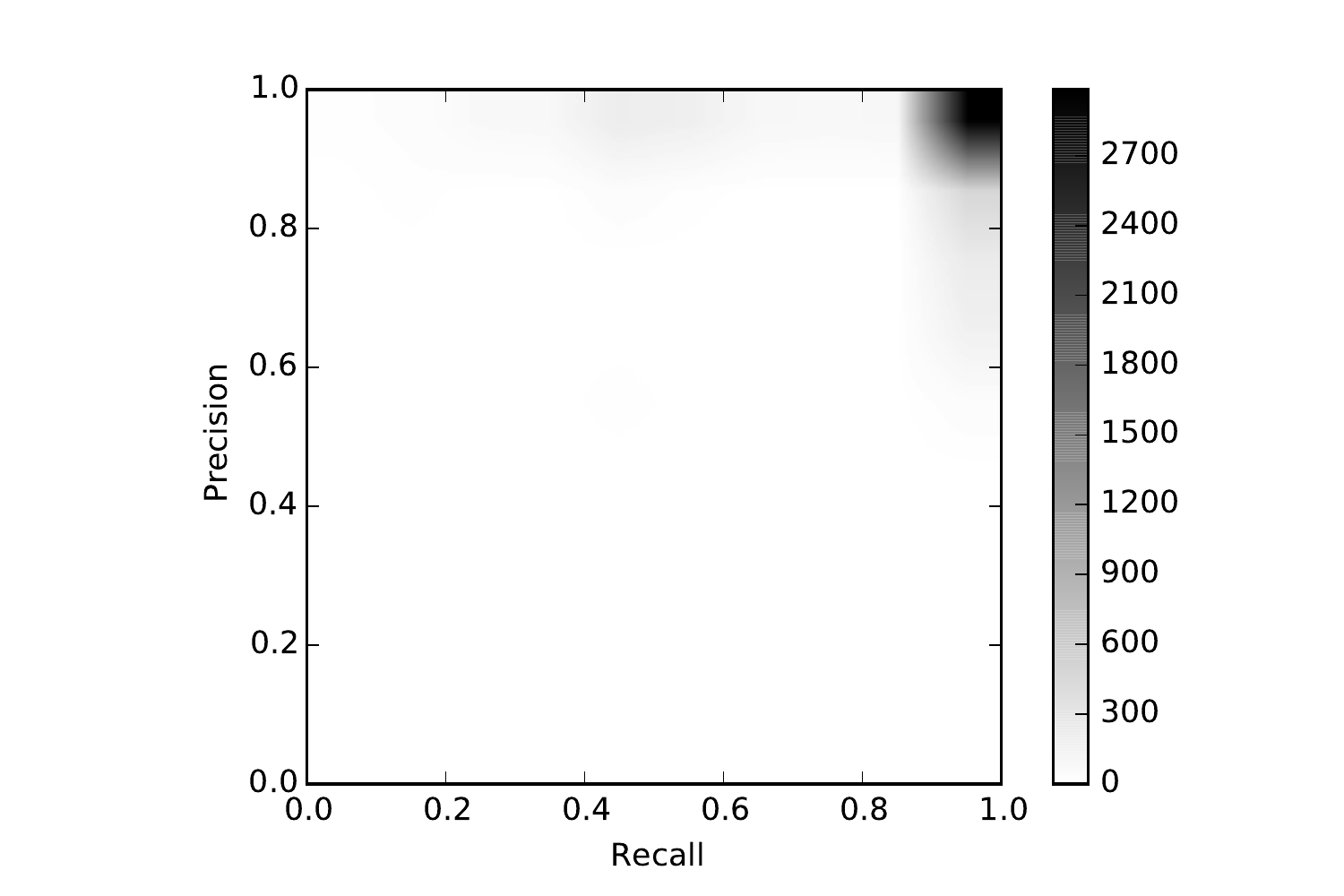}  \\
(a) & (b)  \\ 
\includegraphics[width=0.49\textwidth]{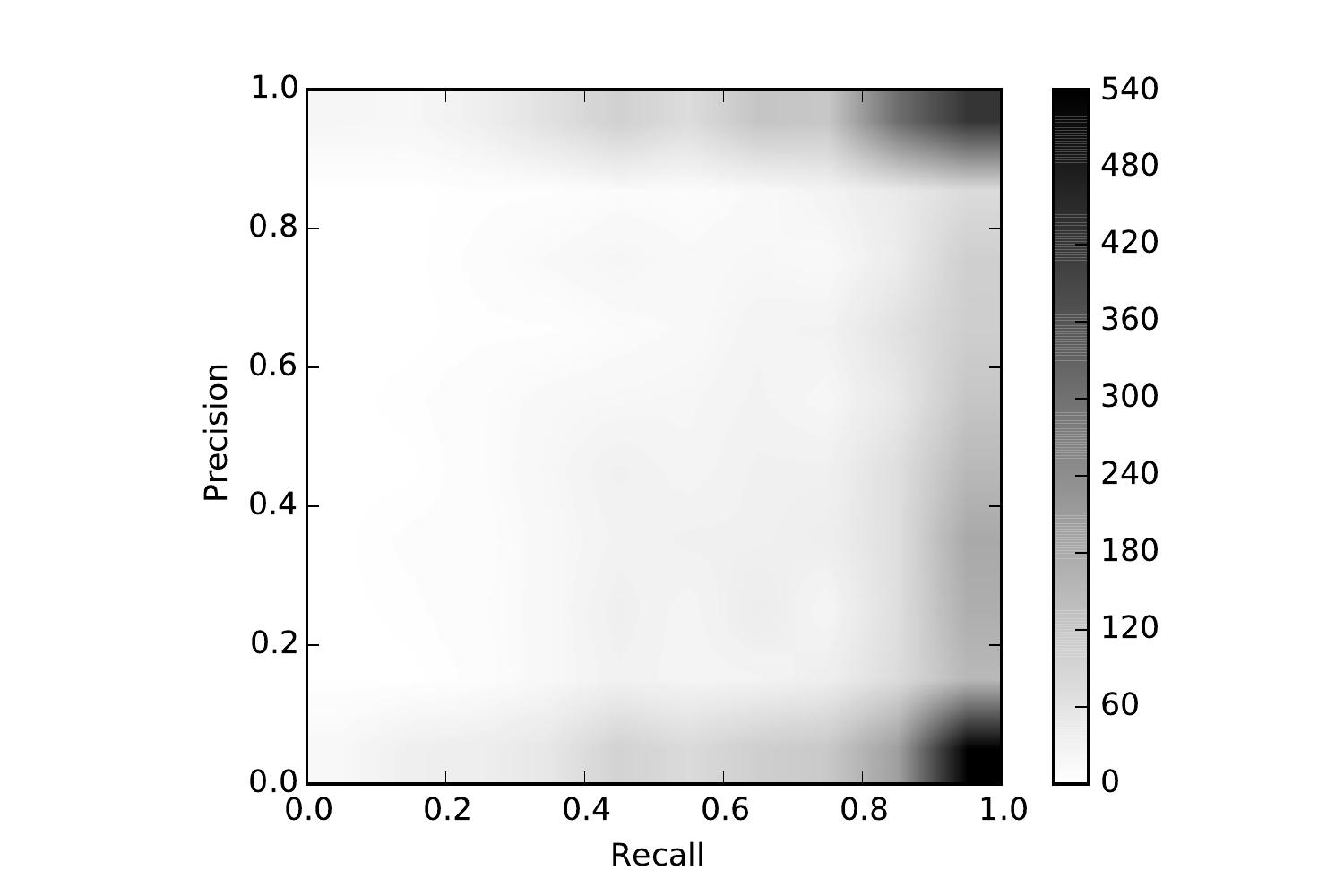} \hspace{3mm} &
\includegraphics[width=0.49\textwidth]{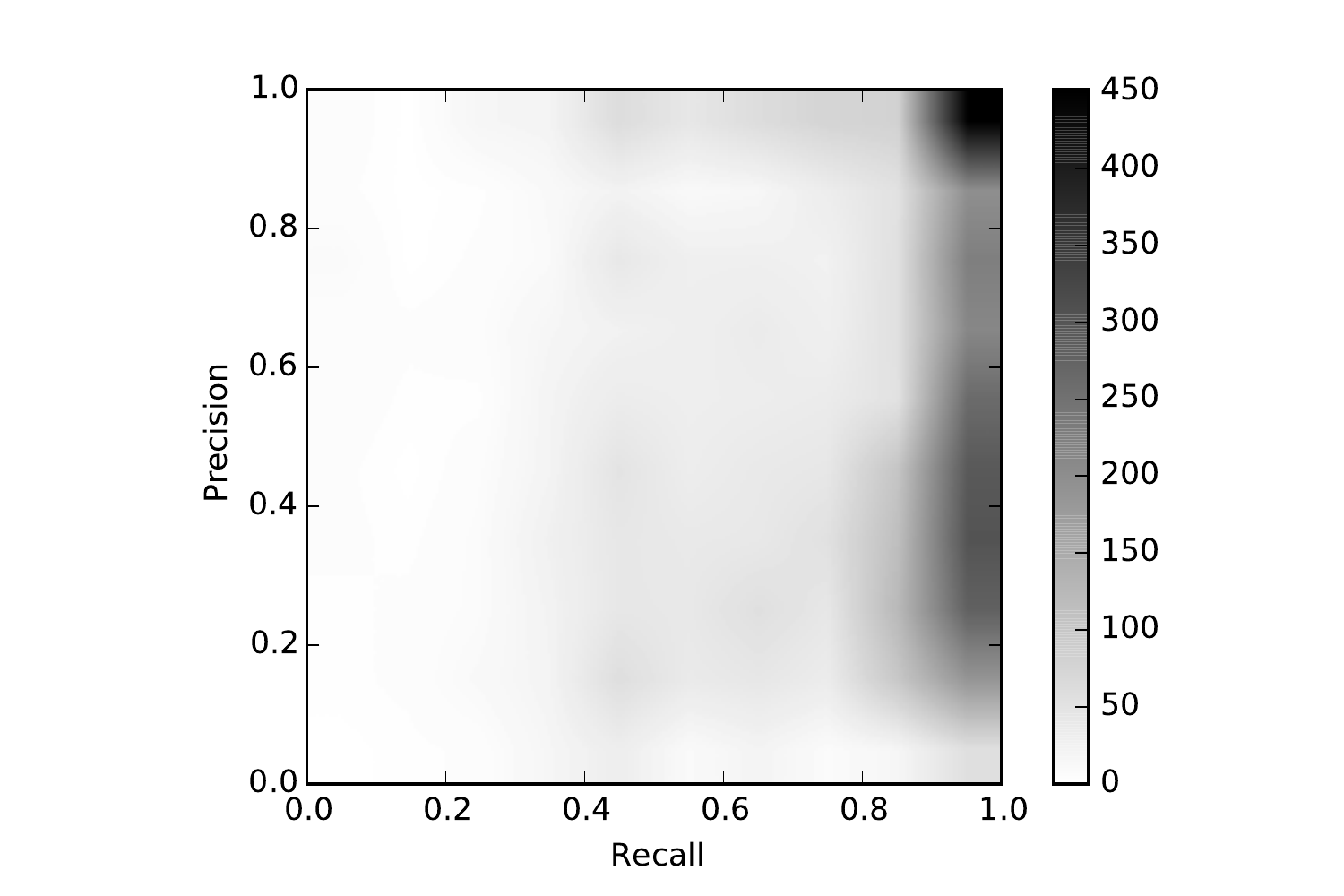}  \\
(c) & (d)  \\
\includegraphics[width=0.49\textwidth]{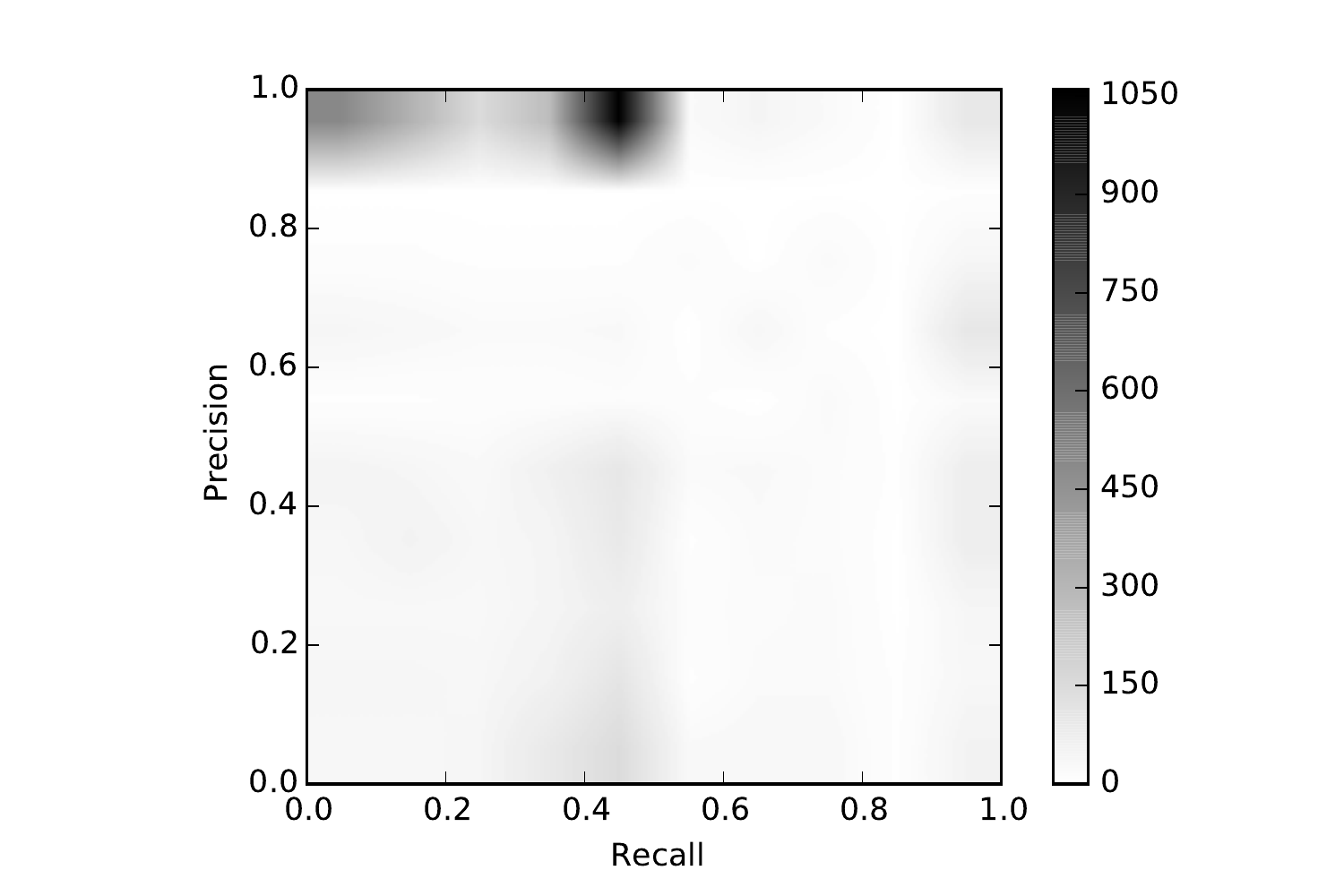} \hspace{3mm} &
\includegraphics[width=0.49\textwidth]{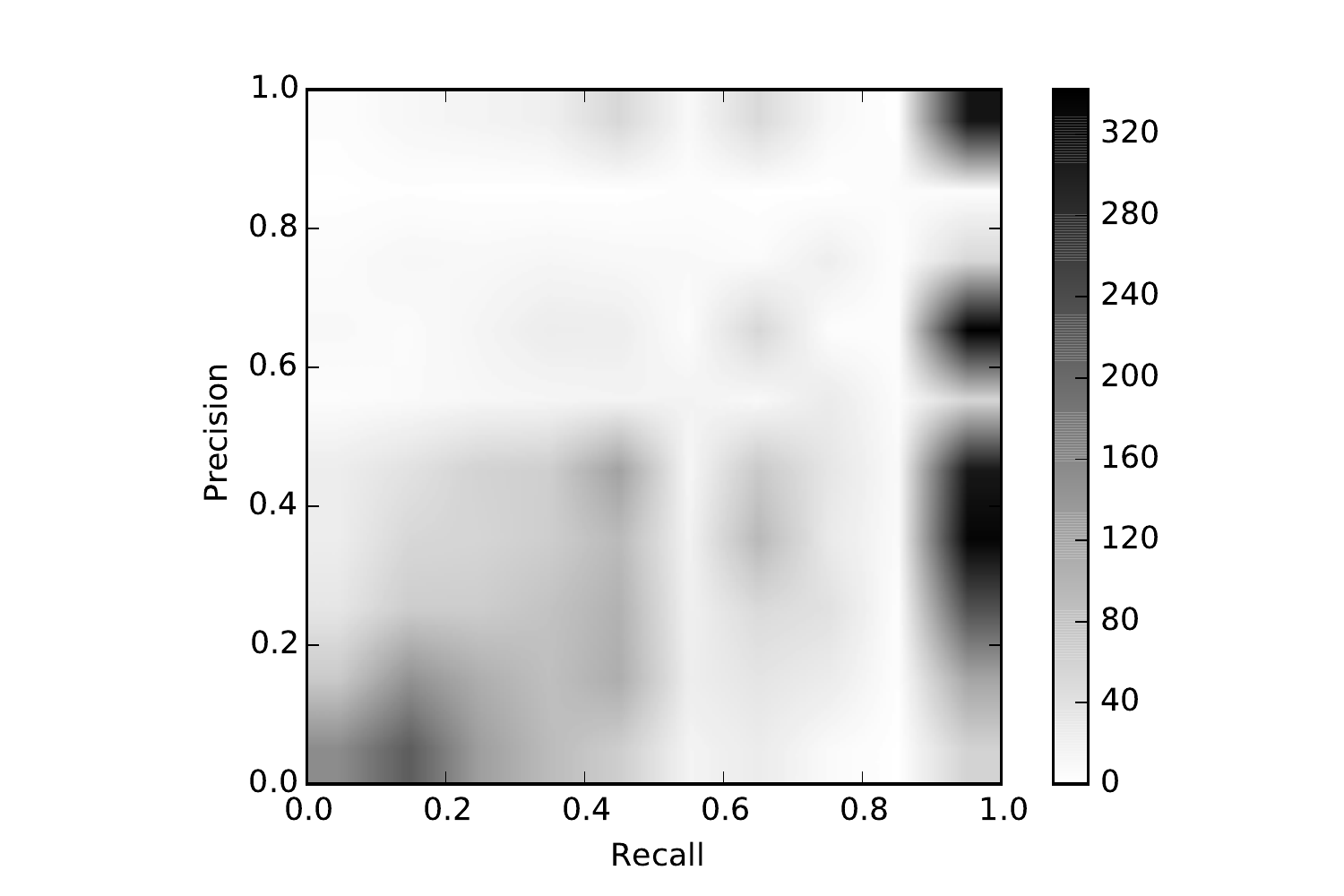}  \\
(e) & (f) 
\end{tabular}
\caption{\label{fig:heatmap}  Precision vs recall heatmap for the (a),(b) Amazon, (c),(d) DBLP, and (e),(f)  Youtube graphs. The first column of plots corresponds to our normalized thresholding method, while the second to   MCL \cite{dongen2000graph}.}
\end{figure*}

TECTONIC provides state of the art performance that can compete
with MCL in terms of quality but is significantly faster.   For instance, on the YouTube graph it is more than 2\,741 times faster than MCL.  Figures~\ref{fig:amazon}(a),(b) show a detailed view of precision and recall as a function of the community size for MCL and our  normalized thresholding method. Figure~\ref{fig:amazon}(c) plots precision vs. recall for our method for various threshold values ranging from 0.01 to 0.1 with a step of 0.01 for all three datasets. Our choice for the threshold in Table~\ref{tab:groundtruth} was the middle choice 0.06.  As the threshold increases,   precision increases and recall decreases.  Finally, it is worth outlining that since many points corresponding to communities in Figures~\ref{fig:amazon}(a),(b) 
fall on the top of each other, we provide   a more detailed view  of recall versus precision in the form of heatmaps, see  Figure~\ref{fig:heatmap}.

Specifically, Figures~\ref{fig:heatmap}(a), \ref{fig:heatmap}(c), and  \ref{fig:heatmap}(e) show the precision and recall for ground-truth communities obtained using our normalized thresholding method for the Amazon, DBLP, and Youtube graphs respectively. Similarly, 
Figures~\ref{fig:heatmap}(b), \ref{fig:heatmap}(d), and  \ref{fig:heatmap}(f) show the precision and recall for ground-truth communities obtained using MCL \cite{dongen2000graph} for the same graphs respectively. These figures  are heatmaps in which darker colors correspond to larger number of communities with given precision-recall tradeoff. In the case of the Amazon graph, MCL's and TECTONIC's outputs resemble each other,  but for the DBLP and Youtube graphs, the two methods produce different outputs that happen to result in comparable precision and recall values.  The figures indicate that while the two methods perform well, in general they behave differently in different regimes.

\section{Theoretical Foundations}
\label{sec:foundations}
\subsection{Preliminaries}

We use a  powerful probabilistic result from \cite{gavinsky2014tail}  to prove our main results in Section~\ref{sec:expanders}.

\begin{definition}[Read-$k$ families]
Let $X_1, \dots,   X_m$ be independent random variables. For $j \in [r]$, let $P_j \subseteq [m]$ and let $f_j$ be a Boolean function of $\{ X_i \}_{i\in P_j}$. Assume that 
$ | \{ j | i \in P_j \} | \leq k$  for every $i\in [m]$. Then, the random variables $Y_j=f_j(\{ X_i \}_{i\in P_j})$ are called a read-$k$ family.
\end{definition}

\begin{theorem}[Concentration of Read-$k$ families] 
\label{thm:readk}

Let $Y_1,\ldots,Y_r$ be a family of read-$k$ indicator variables with $\Prob{Y_i=1}=q$. Also, let $Y=\sum_{i=1}^r Y_i$. Then for any $\epsilon>0$,
\beql{readk-upper-bound}
\Prob{  Y \geq  (1+\epsilon) \Mean{Y} } \leq e^{-\frac{\epsilon^2 \Mean{Y}}{2k(1+\epsilon /3)} }
\eeq  

\beql{readk-lower-bound}
\Prob{  Y \leq  (1- \epsilon) \Mean{Y} } \leq e^{-\frac{\epsilon^2 \Mean{Y}}{2k} }.
\eeq  
\end{theorem}

\subsection{Planted partition model}
\label{planted}

The following example illustrates the benefit of using the triangle biased  walk we described in Section~\ref{conductance} instead of the standard random walk. Let $G \sim G(nk,k,p,q)$ be a graph sampled from 
the planted partition model on $nk$ vertices, with $k$ clusters 
each  with exactly $n$ vertices. Specifically, let $\Psi: V \rightarrow [k]$ be the partition function and let any pair of distinct
vertices $u,v \in V(G)$ connect with probability $p$ if $\Psi(u) = \Psi(v)$  and with probability $q<p$ otherwise.   For the sake of simplicity, assume $p,q$ are two distinct constants.

\begin{lemma}
\label{lem:plantedRW} 
Let $G \sim G(kn,k,p,q)$ be an unweighted graph.  Let $H(V,E,w)$ be the
auxiliary graph derived from $G$ where the graphs edges $(u,v)$ are weighted as $w(u,v) = t(u,v)$, i.e., 
according to the number of triangles that contain edge $(u,v)$. Consider random walks $X_t$ and $Y_t$ on the vertices of $G$ and $H$,
respectively, where the random walk on $G$ is the standard random walk
on the random walk on $H$ a neighbor proportionally to the weights on
the edges.  Then with probability $1-o(1)$ over the choice of $G$, for all vertices
$u$,
$$\Pr(\Psi(X_{t+1}) = \Psi(X_t)~|~X_t = u) < \Pr(\Psi(Y_{t+1}) = \Psi(Y_t)~|~Y_t = u).$$
\end{lemma}

In plain words, Lemma~\ref{lem:plantedRW} shows that the random walk on $H$ is more likely to stay in the same component of the planted partition than the random walk on $G$.  Leveraging these ideas
further, we can show that in the planted partition model,
reweighting edges by triangle counts can completely reveal the cluster structure. 


\subsection*{Proof of Lemma~\ref{lem:plantedRW} }
 
We provide the intuition of the proof by working with expectations;  the full proof
uses relies on concentration of all values around their expectations, which follows from  concentration of measure.

For the random walk on $G$, a vertex $u$ has $p(n-1)$ neighbors in expectation in the same
partition, and $qn$ neighbors in expectation in each other partition.  For simplicity we
use $pn$ as the expectation for the number of neighbors in the same partition as asymptotically
the difference does not matter.  Thus $\Pr(\Psi(X_{t+1}) = \Psi(X_t)~|~X_t = u) = \frac{p}{p+q(k-1)}$ 
with our approximations.  

For the random walk on $H$, we first determine the expected vertex weights.  If $(u,v) \in E(G)$, and
$\Psi(u) \neq \Psi(v)$, then $ \Mean{ w(u,v) } = 2(n-2)pq + (k-2)nq^2$.  The first term corresponds to triangles
where the third vertex is in the same component as $u$ or $v$, the second term to triangle where the
third vertex is in another component.  Similarly, if $\Psi(u) = \Psi(v)$, then 
$  \Mean{ w(u,v) } = (n-2)p^2 + (k-1)nq^2$.  Again for simplicity we avoid lower order terms
and use weights $2npq+(k-2)nq^2$ and $np^2 + (k-1)nq^2$ for the two cases.

For the random walk on $H$, there are in expectation $(n-1)p$ neighbors in the same partition,
and $(k-1)nq$ neighbors in the other partitions.  Hence the total expected weight of edges
to neighbors in the same partition is (again, approximately) $np(np^2 + (k-1)nq^2)$, against
$(k-1)nq(2npq+(k-2)nq^2)$ to other partitions.  We thus find that
 
 \begin{align*} 
\Prob{\Psi(Y_{t+1}) = \Psi(Y_t)~|~Y_t = u } &= 
\frac{p^3+(k-1)pq^2}{p^3+3(k-1)pq^2+(k-1)(k-2)q^3}. 
\end{align*} 

\noindent The following chain of statements are equivalent:

\begin{align*}
\frac{p^3+(k-1)pq^2}{p^3+3(k-1)pq^2+(k-1)(k-2)q^3} > \frac{p}{p+q(k-1)}  \Leftrightarrow & \\
  (k-1)p^3q + (k-1)pq^3 > 2(k-1)p^2q^2   \Leftrightarrow & \\
2pq  < p^2 + q^2.&
\end{align*}

The last statement follows from the arithmetic mean-geometric mean inequality, 
with strict inequality as $p \neq q$.  

The high probability result follows from the fact that all expectations are correct \whp up to lower order terms due to concentration.  Hence with more non-instructive
work we find that \whp for all vertices $u$:

\begin{align*} 
\Prob{ \Psi(X_{t+1}) = \Psi(X_t)~|~X_t = u } =&  \frac{p}{p+q(k-1)} + o(1); 
 \end{align*} 

\begin{align*} 
\Prob{\Psi(Y_{t+1})  = \Psi(Y_t)~|~Y_t = u } =&  
\frac{p^3+(k-1)pq^2}{p^3+3(k-1)pq^2+(k-1)(k-2)q^3} + o(1).  \end{align*}

\noindent The result follows.   $\blacksquare$

We also outline how in the planted partition model
reweighting edges by triangle counts can recover the cluster structure.  (This is a phenomenon observed on real data as well, see Figure~\ref{fig:components1} in Section~\ref{sec:introduction}.)
For example, set $p=\frac{3\log{n}}{\sqrt{n}}, 
q=\frac{\log{n}}{\sqrt{n}}$, and let $G \sim G(2n,2,p,q)$ be a graph sampled according to the planted partition model. The weight of an edge within a cluster $w_{in}$ has expectation ${n-1 \choose 1}p^2+{n \choose 1}q^2\approx  10\log^2{n}$, and similarly the expectation of the weight $w_{out}$
of an edge crossing clusters is $\Mean{w_{out}}=6\log^2{n}$. By Chernoff bounds, we obtain that $\Prob{ w_{in} < 8\log^2{n} } =o(n^{-2})$ 
and similarly $\Prob{ w_{out} > 8\log^2{n} } =o(n^{-2})$.  A union bound over all possible ${n \choose 2}$ edges yields that with high probability all edges within a cluster have
weight at least $8 \log^2{n}$ and all edges crossing clusters have weight at most $8 \log^2{n}$.
It follows immediately that removing edges 
with weight less than $8\log^2{n}$ recovers the two clusters.  A more complete analysis with
bounds on the required ``gap'' between $p$ and $q$ needed to recover clusters
will appear in the full version.

\subsection{Triangle expanders}
\label{sec:expanders}  

We extend the notion of an expander graph to a triangle expander. 
  
\begin{definition} 
A graph $G(V,E)$   is a triangle expander if all subsets $S \subseteq V$ with $|S|=s\leq 0.5 n$  have constant triangle expansion, i.e.,  $\phi_3(S)=\Theta(1)$.
\end{definition} 
We prove that triangle expanders exist.

\begin{theorem} 
\label{thm:thm1}
Let $G \sim G(n,p)$ with $p$ equal to $\frac{\log(n)}{n^{1/3}}$. 
With high probability, $G$ is a triangle expander. 
\end{theorem}

 \noindent Notice that for this range of $p$, the expected number of edges is $O(n^{\tfrac{5}{3}} \log{n})$. An interesting open problem is to show the existence of sparser triangle expanders. We make the following conjecture. 
 
{\bf Conjecture 1:}  $G \sim G(n,p)$ with $p$ equal to $\frac{\log(n)}{n^{2/3}}$ is a triangle expander \whp.

\noindent Also, an interesting question is whether triangle expansion implies edge expansion. Our result is stated as the following theorem.
 
\begin{theorem}
\label{expansion} 
There exist edge expanders that are not triangle expanders. Similarly, under conjecture 1, there exist triangle expanders that are not edge expanders. 
\end{theorem} 

\noindent  Our construction works not only under conjecture 1, but for any triangle expander that has diameter at least 3.

\subsection*{Proof of Theorem~\ref{thm:thm1}}

Consider any cut $(S:\bar{S})$. We prove concentration results
for the number of triangles $t(S:\bar{S})$  cut by $(S:\bar{S})$, and for the triangles induced by $S$ separately. Then, we combine the two concentration results to prove that $\phi_3(G) = \Theta(1)$. 

Define an indicator variable $X_{uv}=\One{ u\sim v}$ for each pair of distinct vertices $u,v \in V$. Notice $\Mean{X_{uv}}=p$.  Let $\epsilon$ be a fixed constant.
 
\noindent  \underline{Number of triangles  $t(S:\bar{S})$ cut by $(S,\bar{S})$.}  
For each value $s=1,\ldots,0.5n$,  define   $Q_s$ to be the event


$$Q_s = \exists S \subseteq V: |S|=s,   \Abs{ t(S:\bar{S}) -  \Mean{t( S:\bar{S} )} }  > \epsilon \Mean{ t(S:\bar{S}) } .$$ 

The random variable $t(S:\bar{S})$ is the sum of two multivariate polynomials, 

$$ t(S:\bar{S}) = \underbrace{\sum\limits_{u \in S, v,w \notin S} X_{uv} X_{vw} X_{uw}}_{T_1(S)}  + \underbrace{\sum\limits_{u,v \in S, w \notin S} X_{uv} X_{vw} X_{uw}}_{T_2(S)}.$$

\noindent The two polynomials are equal to the number of triangles which have exactly one and two vertices in $S$ respectively.  By the independence of the random variables $\{X_{uv}\}$ and the linearity of expectation, $ \Mean{T_1(S)} = {s \choose 1}{n-s \choose 2}p^3$, and 
 $\Mean{T_2(S)} = {s \choose 2}{n-s \choose 1}p^3$. Therefore, 

$$ \Mean{ t(S:\bar{S}) } = \frac{ \log^3(n)}{n} \Bigg( {s \choose 1}{n-s \choose 2} + {s \choose 2}{n-s \choose 1} \Bigg).$$

\noindent We prove  that there exists a constant $c=c(\epsilon)$ such that

$$ \Prob{ \Abs{ t(S:\bar{S}) -  \Mean{t( S:\bar{S} )} }  > \epsilon \Mean{ t(S:\bar{S}) }} \leq e^{- c s \log^3{n} }.$$

\noindent We apply Theorem~\ref{thm:readk}. Here, 
$m={n \choose 2}$, $r= {s \choose 1}{n-s \choose 2} + {s \choose 2}{n-s \choose 1}$. We define the family of variables $Y_{uvw} = X_{uv} X_{vw} X_{uw}$ for each triple of vertices $u,v,w$ such that  either $u \in S, v,w \notin S$ or $u,v \in S, w\notin S$. 
This is a read-$k$ family of variable where $k \leq n$.
We apply Equation~\eqref{readk-upper-bound}

\begin{align*}
 \Prob{ t(S:\bar{S}) \geq (1+\epsilon) \Mean{ t(S:\bar{S}) } } &\leq
\exp\bigg\{ -\frac{ \Mean{t(S:\bar{S})} \epsilon^2}{2nk(1+\epsilon/3)}   \bigg\} \\ 
&\leq \exp\bigg\{  \frac{0.01\epsilon^2 s \log^3{n} }{ 2(1+\epsilon/3)}       \bigg\} \\ 
&= e^{-C(\epsilon) s \log^{3}(n)}
\end{align*} 

\noindent By applying Equation~\eqref{readk-lower-bound} 

\begin{align*}
 \Prob{ t(S:\bar{S}) \leq (1 - \epsilon) \Mean{ t(S:\bar{S}) } } &\leq
 e^{-C'(\epsilon) s \log^{3}(n)},
\end{align*} 

\noindent where $C'(\epsilon)= 0.005 \epsilon^2$.   By taking  two union bounds we get for any constant $\epsilon>0$,

\begin{align*}
\Prob{ Q_s } &\leq {n \choose s} e^{- \min{(C (\epsilon),C' (\epsilon)} ) s \log^3{n} } \leq \big(\frac{en}{s}\big)^s e^{- \min{(C (\epsilon),C' (\epsilon)} s  \log^3{n} } = o(n^{-1}), 
\end{align*}

\noindent and therefore by a union bound, 

$$ \Prob{ \cup_{s=1}^{0.5n} Q_s } \leq n o(n^{-1})=o(1).$$  

\underline{Number of triangles $T_3(S)$ induced by $S$.}
In order to prove that $G \sim G(n,p)$ is a triangle expander 
\whp, it suffices to show that  for all sets 
 $S\subseteq V$, $T_3(S) = O( \Mean{T_1(S)+T_2(S)})$ \whp.  We express $T_3(S)$ as  the multivariate polynomial
$T_3(S) = \sum\limits_{u,v,w \in S } X_{uv} X_{vw} X_{uw}$. Notice that    $\Mean{T_3(S)} = {s \choose 3} p^3$. 

\noindent In the following, we prove that $T_3(S)$ does not exceed twice its expectation \whp. We consider two  cases, depending on the cardinality of the set $S\subseteq V$.

\noindent
\underline{$\bullet$ {\sc Case 1:} $s=o(n)$} 
Consider any fixed set $S \subseteq V$ such that  $|S|=s=o(n)$. For any cardinality $s = o(n)$, we can write $s=\frac{n}{\omega(n)}$ where $\omega(n)$ is an appropriately chosen slowly growing function such that 
$\omega(n) \rightarrow +\infty$ as $n \rightarrow +\infty$.  We obtain 

\begin{align*}
\Prob{ T_3(S) \geq 2 \Mean{ t(S:\bar{S})  }} &\leq e^{- sn\log^2{n} }. 
\end{align*} 

\noindent By taking a union bound over all possible subsets $S \subseteq V, s=o(n)$ we obtain that  

\begin{align*}
\Prob{ \exists S: S\subseteq V, s=o(n), T_3(S) \geq 2\Mean{ t(S:\bar{S})  } } &\leq 
  \sum_{s \leq 0.5n} {n \choose s}  e^{- sn \log^2{n}  } =o(1). 	
\end{align*}

\noindent
\underline{$\bullet$ {\sc Case 2:} $s=\Theta(n)$} \\
Fix any set $S \subseteq V$ such that $s = \alpha n$ for some 
constant $\alpha \leq 0.5$. By applying Equation~\eqref{readk-upper-bound} with $\epsilon=1$  we obtain 

\begin{align*}
\Prob{ T_3(S) \geq 2\Mean{T_3(S)}  }  &\leq
e^{-\tfrac{\epsilon^2}{2(1+\epsilon/3)}   \tfrac{p^3{s \choose 3}}{n}  }   \leq e^{-n\log^2{n} }.
\end{align*}

\noindent  By taking a union bound over all possible subsets $S \subseteq V, s=\Theta(n)$ we obtain that  

\begin{align*}
\Prob{ \exists S: S\subseteq V, s=\Theta(n), t_3(S) \geq2 \Mean{ T_3(S)  } } &\leq \\ 
  \sum_{s=\Theta(n)} {n \choose s}  e^{- n \log^2{n}}      \leq  
 \sum_{s=\Theta(n)}  \bigg(\frac{en}{s} \bigg)^s  e^{- n \log^2{n}} &=o(1).
\end{align*}

\underline{Triangle conductance $\phi_3$.}  By combining our concentration results for $T_3(S), t(S:\bar{S})$, we obtain that \whp for any set $S \subseteq V, |S| \leq 0.5n$  

\begin{align*}
\phi_3(S) &\geq  \frac{ (1-\epsilon) \Mean { t(S:\bar{S}) }}{ 3\times 2\Mean { t(S:\bar{S}) } + 2(1+\epsilon) \Mean { t(S:\bar{S}) } } 
             \geq \frac{2(1-\epsilon)}{7+4\epsilon} = \Theta(1).
\end{align*}

\noindent Therefore, $G \sim G(n, \tfrac{\log{n}}{n^{1/3}})$ is a triangle expander \whp. 

\subsection*{Proof of Theorem~\ref{expansion}}

Since a bipartite network contains no triangles, and there 
exist bipartite expander graphs, the first direction is trivial. Nonetheless, we provide a non-trivial  construction.

(i)  Let $G \sim G(n, \frac{ \log{n} }{n^{1/3}})$.  We modify $G$ 
in such a way that we maintain its edge but not its triangle expansion. 

\underline{Claim 1: Volume is concentrated. }
We prove that for any $S \subseteq V$, $\text{vol}_2(S) \in [(1-\epsilon)\Mean{\text{vol}_2(S)}, (1+\epsilon)\Mean{ \text{vol}_2(S)}]$ \whp.  It suffices to show that for {\em each} vertex $v \in V(G)$, 
$deg(v) \in [(1-\epsilon)\Mean{deg(v)}, (1+\epsilon)\Mean{ deg(v) }]$ \whp. Notice, $deg(v) \sim Bin(n-1, p)$. The claim is easily proved by applying Chernoff and taking a union bound over $n$ vertices. 

\underline{Claim 2: Edges crossing cut are concentrated.}
We prove that for all sets $S \subseteq V$, the number of edges 
$e(S,\bar{S})$ that cross the cut $(S,\bar{S})$ are concentrated around the expectation. First, notice that $e(S,\bar{S}) \sim Bin( s(n-s), p)$.
We define for each possible size $s=1,\ldots,0.5n$  the event 

$$Q_s = \exists S \subseteq V: |S|=s, e(S:\bar{S}) \notin[ (1-\epsilon), (1+\epsilon)] \Mean{ e(S:\bar{S}) } .$$ 

\noindent We apply Chernoff  and union bound.

\begin{align*}
\Prob{ \cup_{s=1}^{0.5n} Q_s }  &\leq 
\sum_{s=1}^{0.5n} {n \choose s} 2e^{-\epsilon^2/3 \tfrac{ s(n-s) \log{n}}{n^{1/3} }} \leq  0.5no(n^{-1}) =o(1). \\
\end{align*}

\underline{Claim 3: Edge conductance is constant whp.} 
By combining claims 1,2 we obtain that for any set $S\subseteq V$ with less than $0.5n$ vertices 

$$\phi_2(S) \geq \frac{ (1-\epsilon) p s(n-s) }{(1+\epsilon) \big(2{s \choose 2} +  s(n-s) \big) } = \Omega(1).$$ 

\noindent Recall that $G$ is also a triangle expander, namely for all sets $S\subseteq V$ with $s \leq 0.5n$ 
$\phi_3(S) = \Theta(1)$. 

Consider  the following modification to $G$. Pick a subset $S$ with $s=n^{2/3}$ vertices and any $X\subseteq S$ with 
$n^{2/3-\gamma}$ vertices, where $\gamma=\frac{1}{10}$. We add a clique 
on $X$ by adding in expectation 
$$ (1- \frac{ \log{n}}{n^{1/3}}) {|X| \choose 2} =(1-o(1)) {n^{\tfrac{2}{3}-\gamma} \choose 2}$$ 

\noindent extra edges. Let $G'$ be the resulting graph. Now, we prove that $G'$ is an edge but not a triangle expander. 

$$\phi_2'(S) \approx  \frac{ pn^{2/3}(n-n^{2/3}) }{ pn^{2/3}(n-n^{2/3}) + 2p {n^{2/3} \choose 2} + {|X| \choose 2} } \rightarrow 1.$$

\noindent It is also easy to check that the conductance of $X$ and any subset of it is constant. For instance,

$$ \phi_2'(X) \approx \frac{ p|X|(n-|X|) }{ {|X| \choose 2} + p|X|(n-|X|) } \rightarrow 1.$$ 

\noindent However, the triangle conductance of $S$ becomes 

\begin{align*}
\phi_3'(S) & \approx  \frac{ p^3  { {s \choose 2} {n-s \choose 1} +  {s \choose 1} {n-s \choose 2} } }{  {|X| \choose 3}+ p^3  { {s \choose 3}+ {s \choose 2} {n-s \choose 1} +  {s \choose 1} {n-s \choose 2} }} =
\frac{ n^{5/3} \log^3{n} }{ n^{2-3\gamma}  +  n^{5/3} \log^3{n} }
=o(1),\\ 
\end{align*}

\noindent since $3\gamma <1/3$.  
 
(ii) We provide a general construction that can be applied to modify any graph that is both an edge and a triangle expander of diameter at least 3 to a graph that is a triangle expander but not an edge expander. 
Notice that $G \sim G(n,p)$ with $p=\frac{\log(n)}{n^{2/3}}$ has diameter at least 3, and under conjecture 1 is a triangle expander \whp. Since the diameter is at least 3, there exists a pair of nodes $u,v$ such $dist(u,v)\geq 3$. We add an edge of arbitrarily large weight between $u,v$. Since  $dist(u,v) \geq 3$, the number of common neighbors $|N(u) \cap N(v)|$ between $u$ and $v$ is 0, so the new edge $(u,v)$ does not change the triangle conductance. However the edge conductance of $\{u,v\}$ becomes arbitrarily close to 0 as we increase the weight of the edge.  $\blacksquare$

\medskip
 
\spara{Motif-based conductance.} The framework we developed for the case of triangles naturally extends to other clique motifs. For instance, if the motif of interest is a clique on four nodes, then we define the $K_4$-conductance $\phi_4$ of a set of nodes $S \subseteq V$ as  $ \phi_4(S) = \frac{ 3c_3 + 4c_4 +3c_1}{12c_4+9c_3+6c_2+3c_1}$,  where $c_i$ is the number of $K_4$ with $i$ nodes in $S$.  Defining appropriate random walks for general motifs, and deriving the conductance in a principled way is an interesting question.  

\section{Conclusion}
\label{sec:concl}

As triangles are a natural indicator of community, we have suggested
formalizing the importance of triangles by considering reweighting edges
according to the number of triangles the edge participates in.  While
our framework is simple, we have shown that it is quite powerful, both
in the more theoretical planted partition model and on real-world
graph experiments.   Another advantage of our approach is that it is amenable to distributed implementations. Furthermore,  it strengthens already existing
approaches based on conductance and spectral clustering.  It also can
generalize naturally to other graph motifs.


Our work suggests several natural open directions.  First, we might
consider variations on the reweighting scheme.  For example, for each
edge in the graph we might use a weight of the form $1+\alpha t(e)$
for some parameter $\alpha$; this way edges would still have some
weight even if they were not part of any triangle.  More generally,
understanding how to set appropriate or approximately optimal edge
weights based on motifs for different applications seems quite
interesting. Also, it is worth exploring the effect of approximate motif counting algorithms, e.g., \cite{kolountzakis2012efficient,pagh2012colorful}, on the clustering performance.    Second, we believe the notion of triangle conductance has further consequences from a theoretical perspective. It would be of interest to better understand its behavior in random graphs, and applications to graph clustering algorithms. Finally, we have not focused on
whether our specific choice of reweighting by triangles might lead to 
especially efficient algorithms designed for this case.

\section*{Acknowledgements} 
The first author thanks Edith Cohen for her feedback. This work was supported in part by NSF grants CNS-1228598, CCF-1320231, and CCF-1535795.


\end{document}